\documentclass[12pt]{amsart}

\usepackage{fullpage,times,url}
\newtheorem{theorem}{Theorem}
\newtheorem{lemma}[theorem]{Lemma}
\newtheorem{corollary}[theorem]{Corollary}

\theoremstyle{definition}
\newtheorem{definition}[theorem]{Definition}

\newcommand{\Linf}{$L_\infty$}
\newcommand{\eps}{\varepsilon}
\newcommand{\Z}{\mathbf{Z}}
\newcommand{\C}{\mathbf{C}}
\newcommand{\CC}{\mathbb{C}}
\newcommand{\half}{\tfrac{1}{2}}
\newcommand{\bull}{\bullet}
\renewcommand{\o}{\otimes}
\renewcommand{\L}{\mathbb{L}}

\begin{document}

\title{Higher derived brackets}

\author{Ezra Getzler}

\begin{abstract}
  We show that there is a sequence of operations on the positively
  graded part of a differential graded algebra $L_\bull$ making it
  into an \Linf-algebra. The formulas for the higher brackets involve
  Bernoulli numbers. The construction generalizes the derived bracket
  for Poisson manifolds, and the Lie 2-algebra associated to a Courant
  algebroid constructed by Roytenberg and Weinstein.
\end{abstract}

\subjclass{17B60, 17B63, 18G55, 53D17}

\keywords{Homotopical algebra, Courant algebroids, differential graded
  Lie algebras, Bernoulli numbers}

\maketitle

The Poisson bracket is a Lie bracket on the vector space $C^\infty(M)$
of functions on a manifold $M$ associated to a Poisson tensor $P$ on
$M$. Koszul \cite{Koszul} showed how the construction of this bracket
could be interpreted in terms of differential graded Lie algebras. The
Poisson tensor induces a differential $\delta_P$ on the graded Lie
algebra $L(M)$ of multivector fields on $M$ (the Schouten algebra):
\begin{equation*}
  \delta_P X = [P,X] : L_i(M) \to L_{i-1}(M) .
\end{equation*}
Here, $L_i(M)$ is the space of smooth sections of the vector bundle
$\Lambda^{1-i}TM$, $-n<i\le1$. The Poisson bracket $\{f,g\}_P$ on
$L_1(M)\cong C^\infty(M)$ is given by the formula
\begin{equation*}
  \{f,g\}_P = [\delta_Pf,g] .
\end{equation*}
In fact, for any differential graded Lie algebra $L$ such that $L_i=0$
if $i>1$, this formula induces a Lie bracket on the vector space
$L_1$.

In this note, we extend this construction, removing the condition that
$L$ vanish above degree~$1$. We show that there is a sequence of
operations on $\L=\tau_{>0}L$ making it into an \Linf-algebra. Note
that we adopt the convention that each of the operations in an
\Linf-algebra lowers degree by $1$. Often, in applications of
\Linf-algebras, the $n$th bracket is taken
to have degree $n-2$; the two conventions differ by a suspension.

\begin{definition}
  An $\boldsymbol{L_\infty}$\textbf{-algebra} is a graded vector space
  $\L_\bull$ with operations
  \begin{equation*}
    \{a_0,\dotsc,a_k\} : \L^{\o k+1} \to \L , \quad k\ge0 ,
  \end{equation*}
  satisfying the following conditions.
  \begin{enumerate}
  \item The operation $\{a_0,\dotsc,a_k\}$ is graded symmetric: for
    all $1\le i\le k$,
    \begin{equation*}
      \label{symmetry}
      \{a_0,\dotsc,a_{i-1},a_i,\dotsc,a_k\} = (-1)^{|a_{i-1}||a_i|}
      \{a_0,\dotsc,a_i,a_{i-1},\dotsc,a_k\} .
    \end{equation*}
  \item the operation $\{a_0,\dotsc,a_k\}$ has degree $-1$:
    \begin{equation*}
      |\{a_0,\dots,a_k\}| = |a_0|+\dots+|a_k|-1 .
    \end{equation*}
  \item For each $n\ge0$, the $\boldsymbol{n}$\textbf{th Jacobi} rule
    holds:
  \begin{equation}
    \label{jacobi}
    \sum_{k=0}^n \sum_{\substack{ I = \{ i_0<\dotsc<i_k \} \\
        J = \{ j_1<\dotsc<j_{n-k} \} \\ I\cup J=\{0,\dotsc,n\} }}
    (-1)^\eps \, \{\{a_{i_0},\dotsc,a_{i_k}\},a_{j_1},\dotsc,a_{j_{n-k}}\} = 0 .
  \end{equation}
  \end{enumerate}
  Here, $(-1)^\eps$ is the sign associated by the Koszul sign
  convention to the action of $\pi$ on the elements $(a_0,\dotsc,a_n)$
  of $\L$.
\end{definition}

Let $\L$ be an \Linf-algebra. By the 0th Jacobi rule
\begin{equation*}
  \{\{a\}\}=0 ,
\end{equation*}
the operation $x\mapsto\{x\}$ is seen to give the graded vector space
$\L$ the structure of a chain complex.

\begin{definition}
  A \textbf{Lie $\boldsymbol{n}$-algebra} is an \Linf-algebra $\L$
  concentrated in degrees $[1,\dots,n]$:
  \begin{equation*}
    0 \longrightarrow \L_n \xrightarrow{\ \delta\ } \dotsc
    \xrightarrow{\ \delta\ } \L_1 \longrightarrow 0 .
  \end{equation*}
  In particular, a Lie $1$-algebra is just a vector space $\L=\L_1$ with
  a Lie bracket $\{a,b\}$.
\end{definition}

\begin{theorem}
  \label{main}
  Let $L$ be a differential graded Lie algebra, with differential
  $\delta$ and bracket $[a,b]$. Let $D$ be the operator on $L$ which
  equals $\delta$ on $L_1$, and vanishes in other degrees. Let $\L$ be
  the positively graded chain complex
  \begin{equation*}
    \L_i =
    \begin{cases}
      L_i , & i>0 , \\
      0 , & i\le 0 .
    \end{cases}
  \end{equation*}
  Then $\L$ is an \Linf-algebra, with brackets
  \begin{equation*}
    \{a\} =
    \begin{cases}
      \delta a , & |a|>1 , \\
      0 , & |a|=1 ,
    \end{cases}
  \end{equation*}
  and, for $n>0$,
  \begin{equation*}
    \{a_0,\dots,a_n\} = b_n \sum_{\pi\in S_{n+1}} (-1)^\eps
    [[\dots[[Da_{\pi_0},a_{\pi_1}],a_{\pi_2}],\dots],a_{\pi_n}] ,
  \end{equation*}
  Here, $(-1)^\eps$ is the sign associated to the action of the
  permutation $\pi$ on the tensor product
  \begin{equation*}
    a_0\o\dots\o a_n\in L^{\o n+1}
  \end{equation*}
  by the Koszul sign convention, and
  \begin{equation*}
    b_n = \frac{(-1)^n\,B_n}{n!} .
  \end{equation*}
  If $L^i=0$ for $i>n$, then $\L$ is a Lie $n$-algebra.
\end{theorem}

We have the following explicit formulas for the first two brackets of
this \Linf-structure:
\begin{align*}
  \{a_0,a_1\} &= \half \Bigl( [Da_0,a_1] - (-1)^{|a_0|} [a_0,Da_1]
  \Bigr) , \\
  \{a_0,a_1,a_2\} &= \tfrac{1}{12} \Bigl( [[Da_0,a_1],a_2] -
  (-1)^{|a_1|} [[a_0,Da_1],a_2] \\
  &\qquad + (-1)^{|a_0|(|a_1|+|a_2|)} [[Da_1,a_2],a_0] -
  (-1)^{|a_0|(|a_1|+|a_2|)+|a_1|} [[a_1,Da_2],a_0] \\
  &\qquad + (-1)^{(|a_0|+|a_1|)|a_2|} [[Da_2,a_0],a_1] -
  (-1)^{(|a_0|+|a_1|)|a_2|+|a_2|} [[a_2,Da_0],a_1] \Bigr) .
\end{align*}

In the case where $L^i$ for $i>2$, this theorem is due to Roytenberg
and Weinstein~\cite{RW}. They formulate their results in the setting
of Courant algebroids, but their approach goes through with no change
at all for any differential graded algebra vanishing above degree
$2$. Our proof in essence generalizes the direct calculations of
Roytenberg and Weinstein.

Bernoulli numbers first arose in the study of differential graded Lie
algebras in the work of Ran \cite{Ran}, whose results have been
considerably clarified by Fiorenza and Manetti \cite{FM}. After the
appearance of an earlier version of this note, D.~Calaque observed to
the author that Theorem \ref{main} is a corollary of the main result
of \cite{FM}. Actually, our methods would also yield a direct proof of
the theorem of Fiorenza and Manetti.

Fiorenza and Manetti prove that if $\phi:K\to L$ is a morphism of
differential graded Lie algebras, then there is a natural
\Linf-structure on the mapping cone $C_\phi[-1]=K[-1]\oplus L$. (Our
conventions for \Linf-algebras differ from theirs by a shift in degree
of $1$; in their paper, they use $C_\phi=K\oplus L[1]$.)

Their construction departs from the differential graded Lie algebra
\begin{equation*}
  \CC_\phi = \{ ( x , f(t) + g(t) \, dt ) \in K\oplus L[t,dt]
  \mid f(0)=0 , f(1)=x \} .
\end{equation*}
Here, $L[t,dt]$ is the module over the free differential graded
algebra $\C[t,dt]$ generated by $t$ and its differential $dt$: that
is, it is the space of $L$-valued differential forms on the affine
line.

There is an inclusion of $C_\phi$ into $\CC_\phi$, which sends
$(x,a)\in K\oplus L[1]$ to $(x,a\,dt)$. There is a chain homotopy $h$
on the complex $\CC_\phi$ which yields a contraction to the subcomplex
$C_\phi$:
\begin{equation*}
  \textstyle
  h \bigl( x , f(t) + g(t) \, dt \bigr) = \bigl( 0 ,
  \int_0^t g(s) \, ds - t \, \int_0^1 g(s) \, ds + x \bigr) .
\end{equation*}
This contraction induces an \Linf-structure on $C_\phi[-1]$ in a
standard way, \emph{via} homological pertubation theory. The brackets
for this structure are sums over binary trees, and a calculation
involving Bernoulli polynomials yields the following explicit
formulas: for $x,y\in K[-1]$ and $a_i\in L$,
\begin{align*}
  \{a\} &= \delta a , \\
  \{x\} &= \phi(x) - \delta x , \\
  \{x,y\} &= (-1)^{|x|} \, [x,y] , \\
  \{x,a_1,\dots,a_n\} &= b_n \sum_{\pi\in S_n} (-1)^\eps
    [[\dots[[x,a_{\pi_1}],a_{\pi_2}],\dots],a_{\pi_n}] ,
\end{align*}
while all other brackets vanish.

Let $\phi$ be the inclusion of $K=\tau_{\le0}L$ into $L$.  There is a
natural quasi-isomorphism from $\L=\tau_{>0}L$ to the mapping cone
$C_\phi[-1]$, which sends $a$ to $(Da,a)$. The \Linf-structure thereby
induced on $\L$ is identical to the one in our theorem.

We now present our direct proof of Theorem \ref{main}, which relies on
the following lemma. The proof is a straightforward application of the
graded Jacobi relation for $L$.
\begin{lemma}
  \label{Z}
  For $j,k\ge0$, and $j+k<n$, consider the expression
  \begin{multline*}
    \Z_{n,j,k} = \sum_{\pi\in S_{n+1}}
    (-1)^{\eps+|a_{\pi_1}|+\dots+|a_{\pi_j}|} \\
    [\dots[[\dots[Da_{\pi_0},a_{\pi_2}],\dots],a_{\pi_j}],
    [\dots[Da_{\pi_{j+1}},a_{\pi_{j+2}}],\dots],a_{\pi_{j+k+1}}]],\dots],a_{\pi_n}]
    ,
  \end{multline*}
  Then $\Z_{n,j,k}=\Z_{n,k,j}$ and if $j+k+1<n$,
  $\Z_{n,j,k}=\Z_{n,j+1,k}+\Z_{n,j,k+1}$.
\end{lemma}

\begin{corollary}
  \label{f}
  The expression
  \begin{equation*}
    F = \sum_{i,j} a_{ij} \Z_{n,i,j}
  \end{equation*}
  vanishes if $f(s,t)+f(t,s)$ lies in the ideal generated by $s+t=1$,
  where $f$ is the polynomial in two variables
  \begin{equation*}
    f(s,t) = \sum_{i,j} a_{ij} s^i t^j \in \mathbb{C}[s,t] .
  \end{equation*}
\end{corollary}

\begin{proof}[Proof (of Theorem \ref{main})]
  The cases $n=0$ and $n=1$ of the Jacobi rule Eq.\ \eqref{jacobi} are
  easily checked directly, so from now on, we assume that $n>1$.

  The contribution of the terms with $k=0$ and $k=n$ to the $n$th
  Jacobi rule is
  \begin{multline*}
    \{\{a_0,\dots,a_n\}\} + \sum_{i=0}^n
    (-1)^{|a_0|+\dots+|a_{i-1}|}
    \, \{a_0,\dots,\{a_i\},\dots,a_n\} \\
    \begin{aligned}
      &= b_n \sum_{\pi\in S_{n+1}} (-1)^\eps \Bigl\{
      \delta[\dots[Da_{\pi_0},a_{\pi_1}],\dots],a_{\pi_n}] \\
      & \qquad - \sum_{i=1}^n (-1)^{|a_{\pi_1}|+\dots+|a_{\pi_{i-1}}|}
      \,
      [\dots[Da_{\pi_0},a_{\pi_1}],\dots],(\delta-D)a_{\pi_i}],\dots],a_{\pi_n}]
      \Bigr\} \\
      &= b_n \sum_{i=0}^{n-1} \Z_{n,i,0} .
    \end{aligned}
  \end{multline*}
  Here, we have used that $Da_{\pi_0}$ vanishes unless
  $|a_{\pi_0}|=1$: this is the source of the minus sign on the third
  line.

  Next, we calculate the contribution of the terms with $k=1$:
  \begin{multline*}
    \sum_{\substack{ I = \{ i_0<i_1 \} \\
        J = \{ j_1<\dotsc<j_{n-1} \} \\ I\cup J=\{0,\dotsc,n\} }}
    (-1)^\eps \, \{\{a_{i_0},a_{i_1}\},a_{j_1},\dotsc,a_{j_{n-1}}\} \\
    \begin{aligned}
      &= b_1b_{n-1} \left( \sum_{\pi\in S_{n+1}} (-1)^\eps
        [\dots[D[Da_{\pi_0},a_{\pi_1}],a_{\pi_2}],\dots],a_{\pi_n}]
        - \sum_{i=0}^{n-2} \Z_{n,i,1} \right) \\
      &= b_1b_{n-1} \Bigl( \Z_{n,0,0} - \sum_{i=0}^{n-2} \Z_{n,i,1} \Bigr) .
    \end{aligned}
  \end{multline*}

  When $n=2$, we see that the Jacobi identity becomes
  \begin{equation*}
    b_2 \Z_{2,0,0} + b_2 \Z_{2,1,0} + b_1^2 \Z_{2,0,0} - b_1^2
    \Z_{2,0,1} = 0 .
  \end{equation*}
  Here, $b_2=\tfrac{1}{12}$, and $b_1=\half$, while
  $\Z_{2,0,0}=2\Z_{2,1,0}$ by Lemma \ref{Z}, and so the whole
  expression does indeed sum to $0$.

  At last, we calculate the contribution of the terms with $1<k<n$ to
  Eq.\ \eqref{Jacobi}:
  \begin{align*}
    \sum_{\substack{ I = \{ i_0<\dotsc<i_k \} \\
        J = \{ j_1<\dotsc<j_{n-k} \} \\ I\cup J=\{0,\dotsc,n\} }}
    (-1)^\eps \, \{\{a_{i_0},\dots,a_{i_k}\},a_{j_1},\dotsc,a_{j_{n-k}}\}
    &= - b_k \, b_{n-k} \sum_{j=0}^{n-k} \Z_{n,j,k-1} \\
    &= b_k \, b_{n-k} \bigl( \Z_{n,n-k,k-1} - \Z_{n,0,k-1} \bigr) .
  \end{align*}
  
  When $n$ is odd, only $k=1$ and $k=n-1$ contribute to the Jacobi
  identity, which becomes
  \begin{equation*}
    b_1 b_{n-1} \Bigl( \Z_{n,0,0} - \sum_{i=0}^{n-3} \Z_{n,i,1} -
    \Z_{n,0,n-2} \Bigr) = 0 .
  \end{equation*}
  This identity holds by Corollary \ref{f}:
  \begin{align*}
    f(s,t) &= b_1 b_{n-1} \bigl( 1 - \sum_{i=0}^{n-3}
    s^it - t^{n-2} \bigr) \\
    &= 1 - \left( \frac{1-s^{n-2}}{1-s} \right) t - t^{n-2} .
  \end{align*}
  This polynomial is congruent to $s^{n-2}-t^{n-2}$ modulo $s+t-1$,
  and hence satisfies the necessary condition of Corollary~\ref{f}.

  When $n>2$ is even, the identity becomes
  \begin{equation*}
    b_n \sum_{i=0}^{n-1} \Z_{n,i,0} +
    \sum_{k=2}^{n-2} b_kb_{n-k} \bigl( \Z_{n,0,k-1} - \Z_{n,n-k,k-1}
    \bigr) = 0 .
  \end{equation*}
  Replacing $\Z_{n,i,j}$ by $s^it^j$, we obtain the power series
  \begin{equation*}
    b_n \sum_{i=0}^{n-1} s^i + \sum_{k=2}^{n-2} b_k b_{n-k} \bigl(
    t^{k-1} - s^{n-k}t^{k-1} \bigr) .
  \end{equation*}
  Modulo $s+t-1$, this is congruent to
  \begin{equation*}
    \sum_{k=0}^n b_kb_{n-k} (1-s^{n-k})t^{k-1} .    
  \end{equation*}
  Multiplying by $x^n$ and summing over $n$, we obtain
  \begin{equation*}
    A(s,t,x) = (f(x)-f(sx))f(tx)/t ,
  \end{equation*}
  where
  \begin{equation*}
    f(x) = 1 + \sum_{n=2}^\infty \frac{B_n\,x^n}{n!} =
    \frac{x/2}{\tanh(x/2)} = \frac{x}{e^x-1} + \frac{x}{2} .
  \end{equation*}
  By Corollary \ref{f}, we see that the identity will follow if
  $A(s,t,x)+A(t,s,x)$, lies in the ideal generated by $s+t-1$ in
  $(st)^{-1}\mathbb{Q}[s,t][\![x]\!]$ up to terms of degree $2$ in
  $x$. But
  \begin{equation*}
    A(s,t,x) + A(t,s,x) = \frac{x^2}{4} \left\{ 1 +
    \frac{4e^x(e^{(s+t-1)x}-1)}{(e^x-1)(e^{sx}-1)(e^{tx}-1)} 
    - (s+t-1) \frac{(e^{sx}+1)(e^{tx}+1)}{(e^{sx}-1)(e^{tx}-1)}
  \right\} ,
  \end{equation*}
  proving the theorem.
\end{proof}


\begin{thebibliography}{9}

\bibitem{FM} Fiorenza, D., and Manetti, M.: \emph{\Linf structures on
    mapping cones,} Algebra Number Theory \textbf{1}, 301–-330
  (2007). \url{arXiv:math/0601312}

\bibitem{Koszul} Koszul,J.-L.: \emph{Crochet de Schouten-Nijenhuis et
    cohomologie,} In: ``The mathematical heritage of \'Elie Cartan
  (Lyon, 1984),'' Ast\'erisque, Numero Hors Serie, 257–-271 (1985).

\bibitem{Ran} Ran, Z.: \emph{Lie atoms and their deformations,}
  Geom. Funct. Anal. \text{18}, 184-–221, (2008).
  \url{arXiv:math/0412204}

\bibitem{RW} Roytenberg, D., and Weinstein, A.: \emph{Courant
    algebroids and strongly homotopy Lie algebras,}
  Lett. Math. Phys. \textbf{46}, 81-–93
  (1998). \url{arXiv:math/9802118}

\end{thebibliography}
\end{document}